\documentclass{PoS}

\title{Vector Boson Fusion Production of the Standard Model Higgs at the LHC}

\ShortTitle{VBF Higgs Production at the LHC}

\author{\speaker{M\'onica Luisa V\'azquez Acosta (on behalf of the CMS Collaboration)}\\
        Imperial College London\\
        E-mail: \email{monicava@mail.cern.ch}}

\abstract{The cross section measurements of the Higgs boson production in the vector boson fusion (VBF) process at the LHC followed by a Higgs boson decay into $\tau \tau$, $WW$ and $\gamma \gamma$ will significantly extend the possibility of Higgs boson coupling measurements. Prospective analyses with the CMS experiment are discussed for the $H \rightarrow \gamma\gamma$, $WW$ and $\tau\tau$ decay channels for an integrated LHC luminosity of 30\ fb$^{-1}$. For a Higgs boson mass in the range 115 to 140 GeV, an observation with a significance above 2 standard deviations is expected in the H to $\gamma\gamma$ channel, and above 3 standard deviations in the H to $\tau\tau$ channel. The H to WW channel offers a discovery reach above 5 sigma in the mass range 140 to 200 GeV. A new complete strategy is presented for the control of systematics and early searches at very low luminosities of the order of 1 fb$^{-1}$.
}

\FullConference{2008 Physics at LHC\\
		 September 29 - 4 October 2008\\
		 Split, Croatia}

\begin{document}

\section{Introduction}
Vector Boson Fusion (VBF) Higgs boson production is the second largest production mechanism at the
LHC. The cross section measurements of the VBF process, 
$VV \rightarrow H$ ($qq \rightarrow qqH$), followed by Higgs boson
decays into $\tau \tau$, $WW$ and $\gamma \gamma$ will significantly extend the
possibility of Higgs boson coupling measurements~\cite{Zeppenfeld:2000td,Duhrssen:2004cv}.

\begin{figure}[b]
\vspace{-3mm}
\begin{minipage}{6cm}
\hspace{1.cm}
 \resizebox{5.cm}{!}{\includegraphics{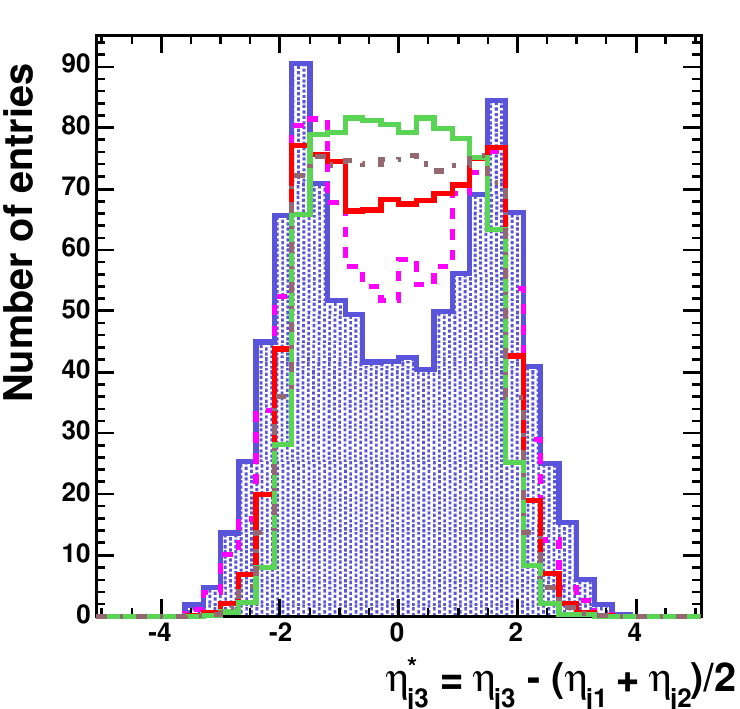}}
\end{minipage}
\begin{minipage}{2.cm}
 \resizebox{3cm}{!}{\includegraphics{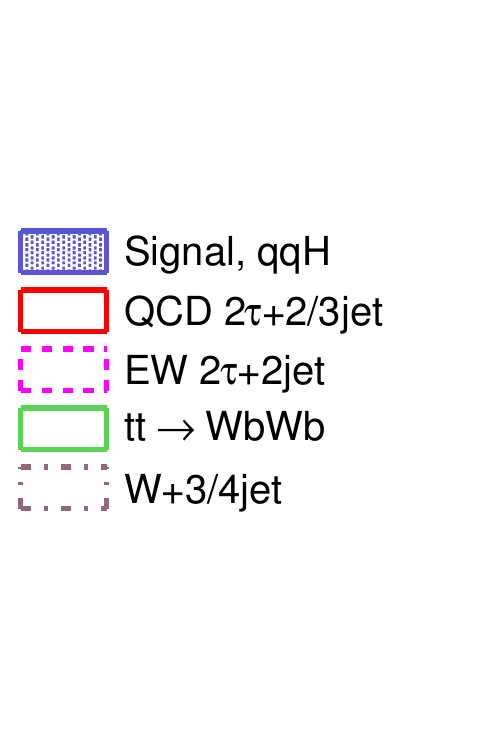}}
\end{minipage}
\begin{minipage}{10cm}
\hspace{-.2cm}
 \resizebox{7.cm}{!}{\includegraphics{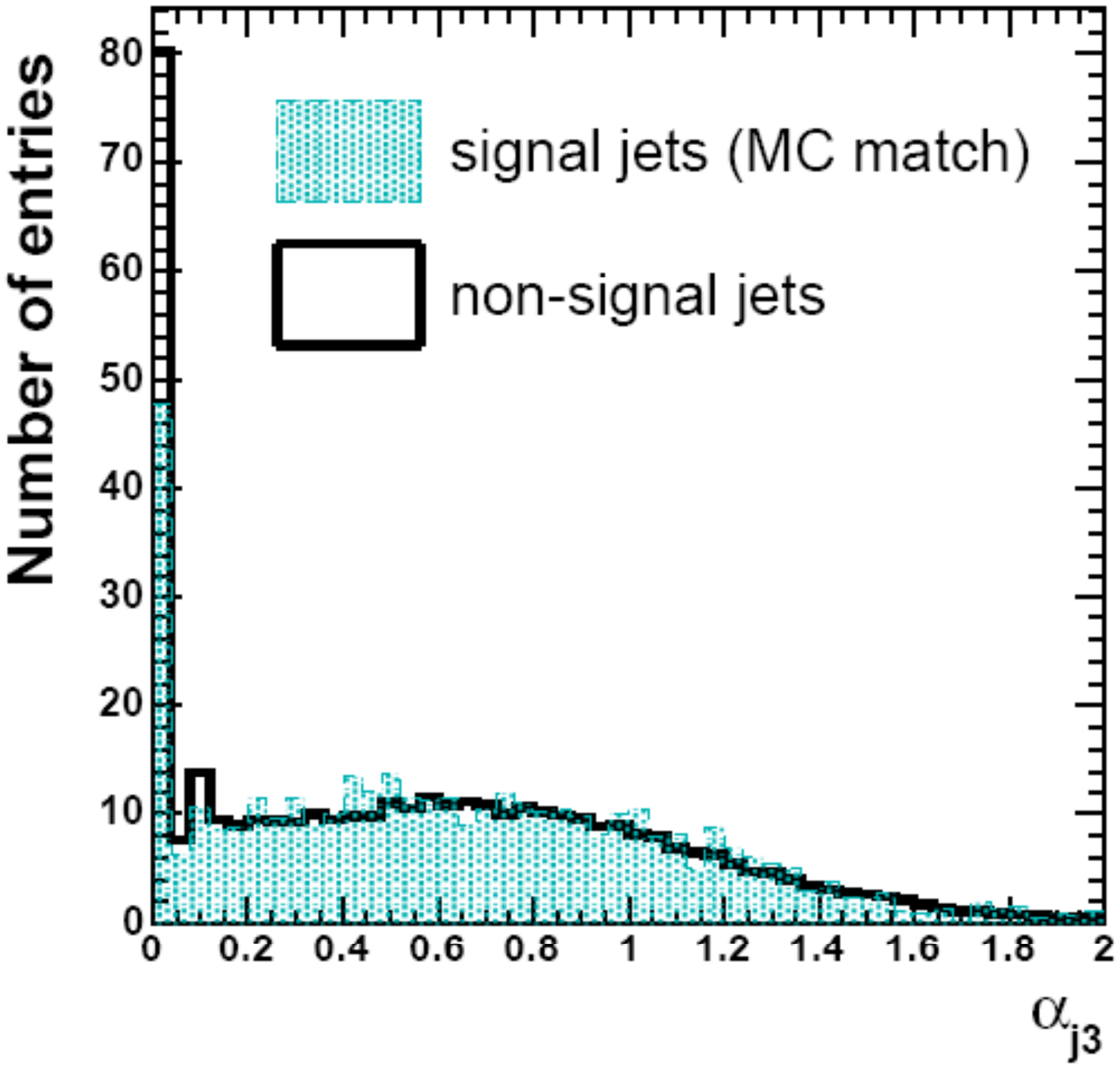}}
\end{minipage} \\
\caption{The $\eta$ distribution of the 3rd jet with respect to the two forward jets (left). The distribution of  $\alpha_{\rm 3j}$ which is used to match jets to the signal vertex (right). }
\label{fig:eta0_alpha}
\vspace{5mm}
\end{figure}

\section{Vector Boson Fusion Signature}

\begin{figure}
\includegraphics[width=.45\textwidth]{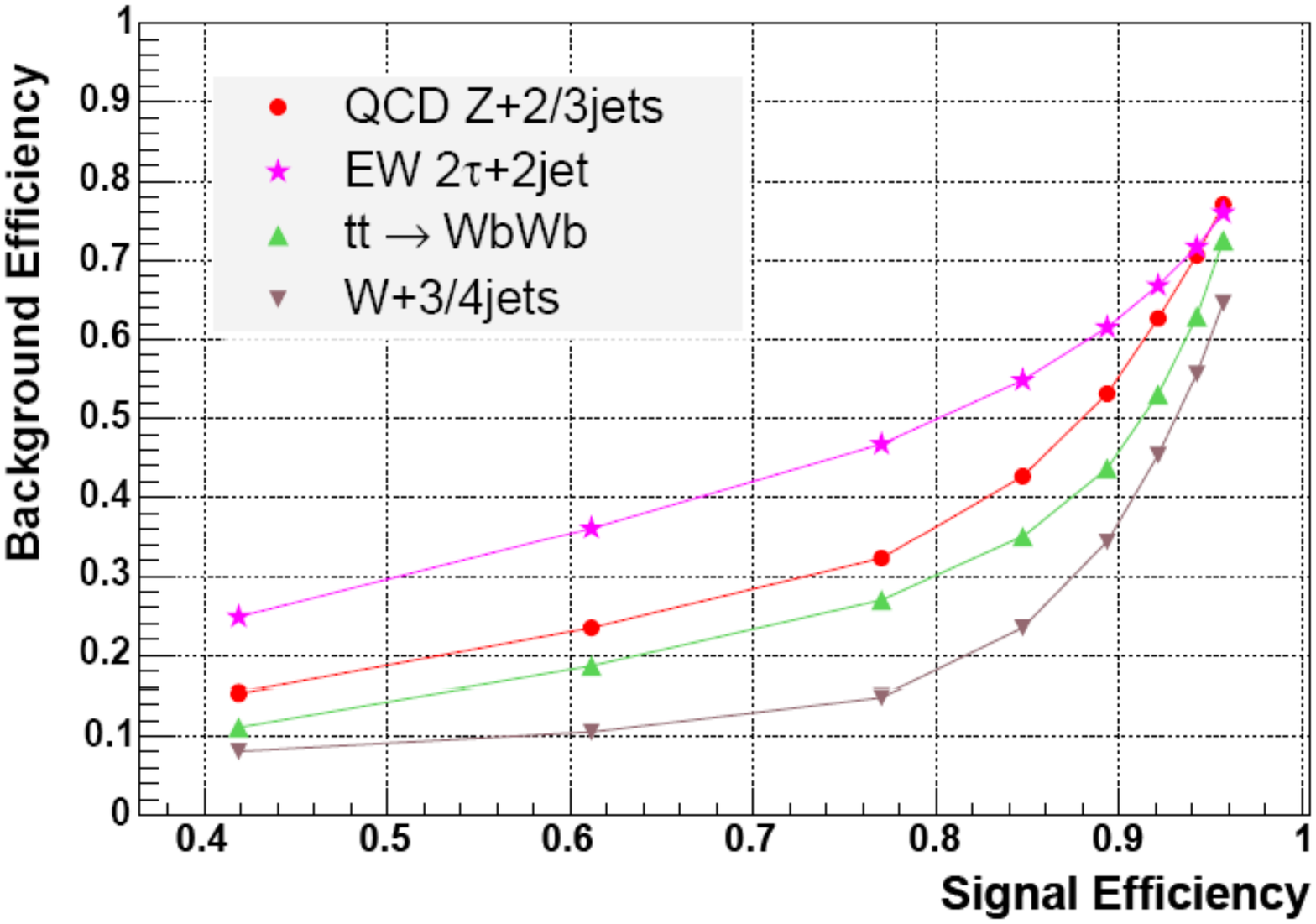}
\includegraphics[width=.3\textwidth]{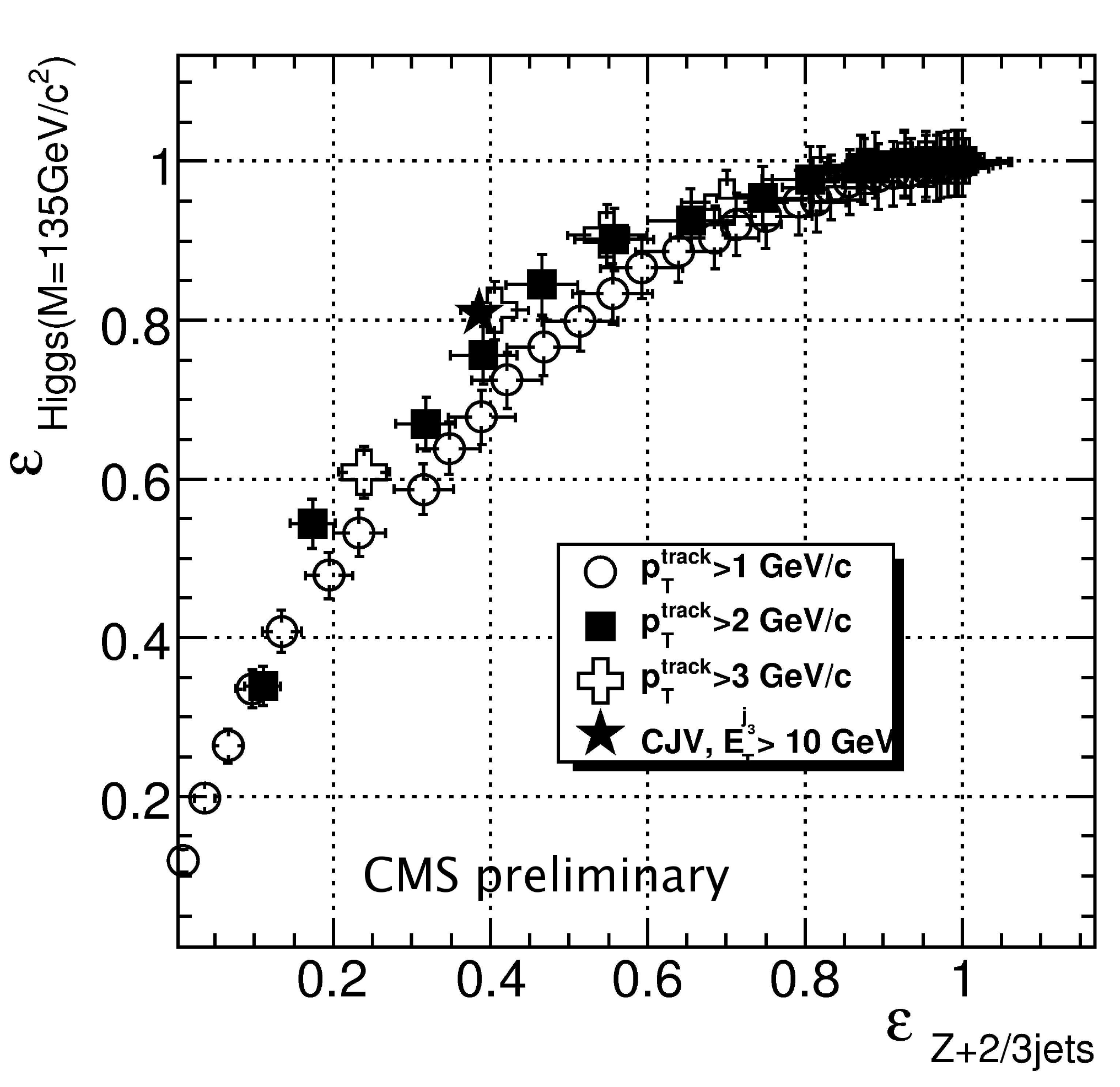}
\caption{Efficiency of the CJV for background versus signal (M$_{\rm H}$=135 GeV), for increasing 3rd jet $\rm E_{Tth}$ threshold (left). 
 TCV performance for different $p_{t}^{track}$ and track multiplicity thresholds compared to the performance of the CJV.}
\label{fig:jetveto}
\end{figure}

Events produced by VBF are characterized by a distict topology of the final
state: two forward jets with little extra hadronic activity and the decay products
of the Higgs boson. The rapidity distribution of the 3rd jet with respect to the two
forward jets, $\eta^{*}_{\rm{j3}}$, is shown in Fig.~\ref{fig:eta0_alpha} (left) 
which shows a double-peak structure for
the electroweak processes, including the VBF signal, and is more central for the
 QCD background samples. Applying a central jet veto (CJV) is a 
poweful rejection method against the QCD background. To avoid considering jets from pile-up events in the CJV,
jets are associated to the signal vertex using tracks. 
For every extra jet one can define the quantity $\alpha_{\rm j3} = 
\Sigma\rm p_{Ttrk} / \rm E_{Tj3}$, where $\rm p_{Ttrk}$ is the $\rm p_{T}$ 
of tracks from the signal vertex within the jet cone and 
$\rm E_{Tj3}$ is the jet measured raw $\rm E_{T}$.  
Figure~\ref{fig:eta0_alpha} (right) shows 
$\alpha_{\rm 3j}$ 
tends to peak at 
low values for non-signal jets.
The efficiency of the veto for the background samples versus the signal efficiency 
 is shown in Figure~\ref{fig:jetveto} (left) for events containg a 3rd jet with $E_{T}$
larger than different threshold values.
An optimal threshold 
where the signal process has $\sim$80\% 
efficiency while the backgrounds are suppressed below 50\% is used~\cite{CMS_NOTE_2006-088}.
An alternative approach 
is to consider a track counting veto (TCV)~\cite{PAS_HIG_08_001},
where the number of
tracks between the two leading jets is counted with different
$p_{T}$ thresholds. 
Figure~\ref{fig:jetveto} (right) shows the performance of the TCV algorithm, i.e the efficiency 
of selecting the signal versus the background for events with an increasing cut on the track 
multiplicity and $p_{T}$. The black star indicates the performance of the 
CJV based on calorimeter jets. The TCV algorithm can reach similar discrimination power than the central jet veto.

\section{Vector Boson Fusion Higgs Discovery Potential}
\begin{figure}[h]
\begin{minipage}[c]{.4\linewidth}
\vspace{-1.7cm}
\hspace{.7cm}
\includegraphics[width =.8\textwidth]{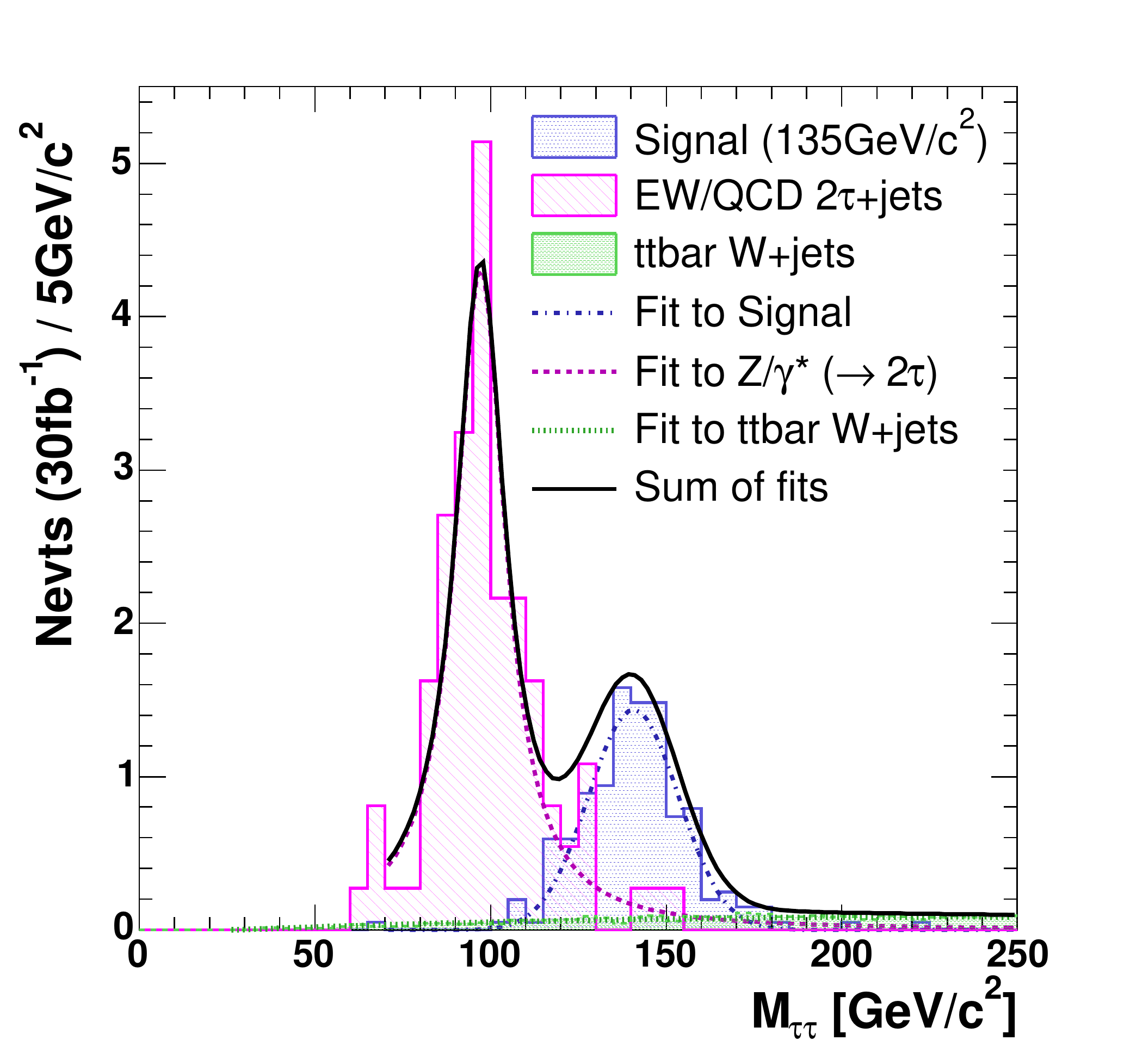}
\end{minipage}
\begin{minipage}[b]{.6\linewidth}
\vspace{+0.9cm}
{\small
\begin{tabular}{|l||c|c|c|c|}
\hline
M$_H$ [GeV] & 115 & 125 & 135 & 145  \\
\hline
\hline
N$_{\rm S}$ (30 fb$^{-1}$) & 10.47 & 7.79 & 7.94 & 3.63 \\
\hline
N$_{\rm B}$ (30 fb$^{-1}$) & 3.70 & 2.21 & 1.84 & 1.42 \\
\hline
S$_{cP}$ at 30 fb$^{-1}$ (no uncertainty) & 4.04 & 3.71 & 3.98 & 2.19 \\
S$_{cP}$ at 30 fb$^{-1}$ ($\sigma_B=7.8\%$) & 3.97 & 3.67 & 3.94 & 2.18 \\
\hline
S$_{cP}$ at 60 fb$^{-1}$ ($\sigma_B=5.9\%$) & 5.67 & 5.26 & 5.64 & 3.19 \\
\hline
\end{tabular}
}
\end{minipage}
\caption{Di-$\tau$ invariant mass expected for a luminosity of 30 fb$^{-1}$ (left). Significance of the expected number of signal events for different Higgs boson masses (right).}
\label{fig:ditau_ptdr}
\end{figure}

The observability of the VBF Higgs boson production has been studied
with the full CMS detector simulation in the
$H \rightarrow \tau\tau$, $\gamma\gamma$ and $WW$ decay channels~\cite{PTDRvol2}.
VBF $H \rightarrow \tau\tau$ production has been studied in the Higgs mass range of 115 to
145 GeV in the lepton plus $\tau_{\rm jet}$ final state. Figure~\ref{fig:ditau_ptdr} (left)
shows the expected di-$\tau$ mass distribution using the collinear approximation~\cite{CMS_NOTE_2006-088}
for a luminosity of 30 fb$^{-1}$. Figure~\ref{fig:ditau_ptdr} (right) shows the significance
of the expected number of signal events for different Higgs masses. A statistical signal significance of 3.9$\sigma$ is
expected for a Higgs mass of 135 GeV.

\vspace{-1.cm}

\begin{figure}[t]
\begin{center}
\includegraphics[width=.28\textwidth]{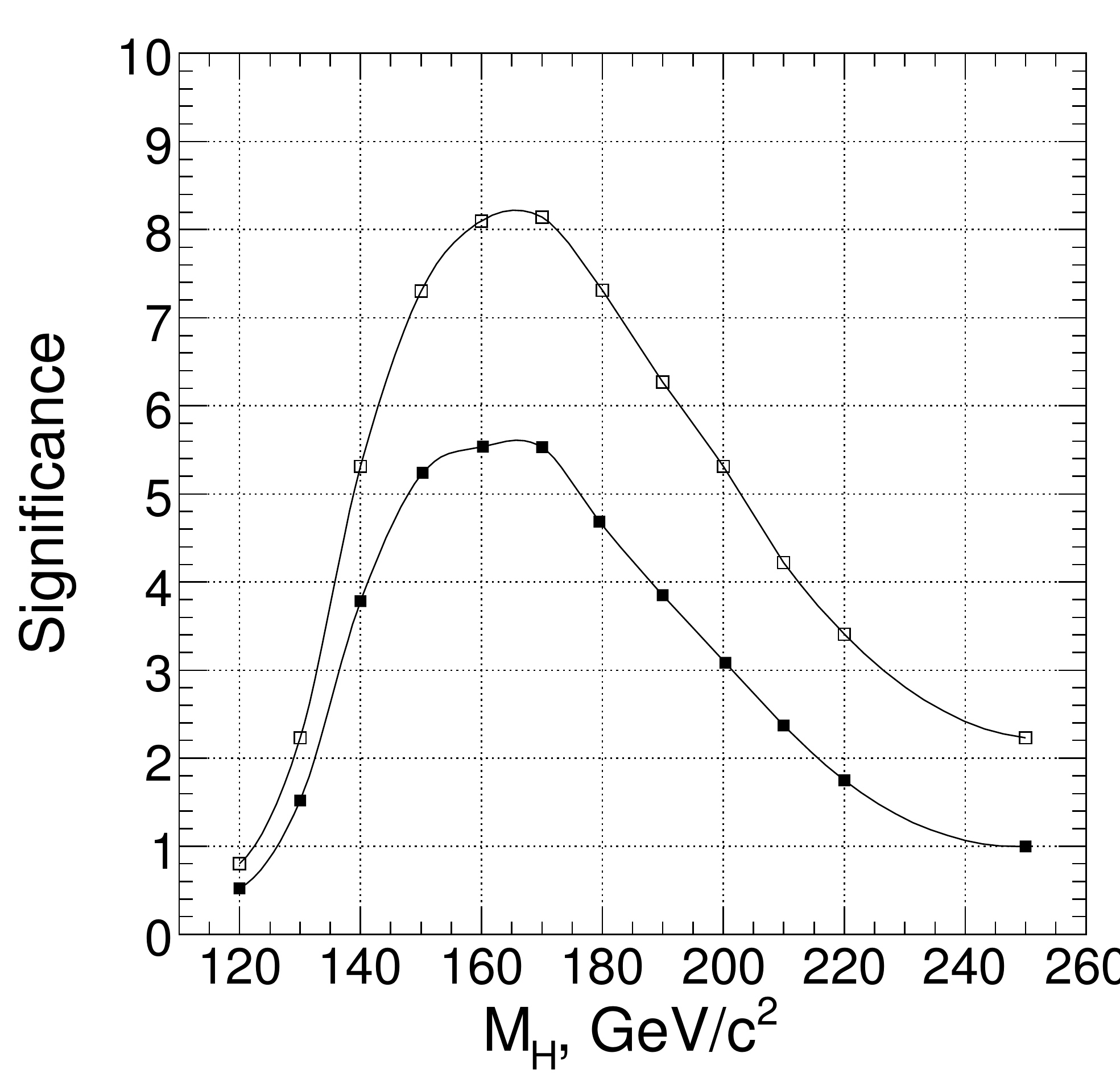}
\includegraphics[width=.38\textwidth]{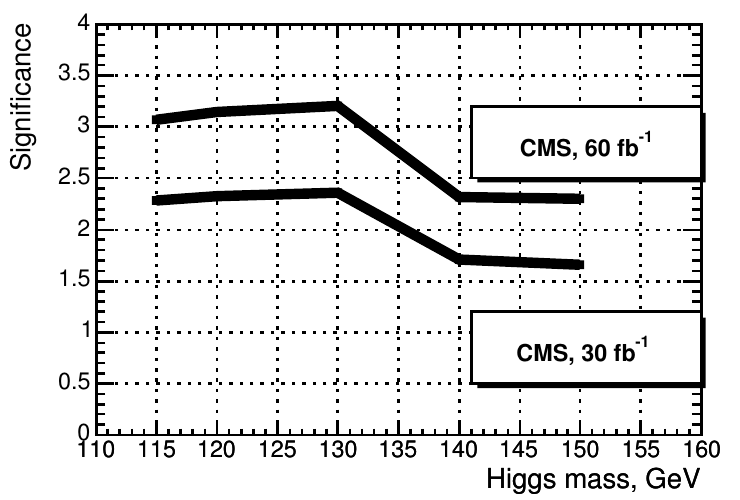}
\caption{Signal significance of VBF $H\rightarrow WW$ for 30 fb$^{-1}$. The high (low) curves correspond to full (loose) extra jet veto (left). Signal significance of VBF $H\rightarrow \gamma\gamma$ for 30 and 60 fb$^{-1}$ (right).}
\label{fig:sig_w_gamma}
\end{center}
\end{figure}

VBF $H \rightarrow WW$ production in the lepton plus two jet final state 
has been studied in the Higgs mass range between 120 and 250 GeV. 
Figure~\ref{fig:sig_w_gamma} (left) shows the signal significance expected with 30 fb$^{-1}$
for different central jet veto selections~\cite{CMS NOTE-2006/092}. In the mass range
between 140-200 GeV a 5$\sigma$ significance can be achieved.
VBF $H \rightarrow \gamma\gamma$ production has also been studied in the Higgs mass range
between 115 and 150 GeV~\cite{CMS NOTE-2006/097}. Figure~\ref{fig:sig_w_gamma} (right) shows the 
signal significance expected with 30 and 60 fb$^{-1}$. With 60 fb$^{-1}$ of collected data 
a 3$\sigma$ significance can be achieved for a low mass Higgs in the range 115 to 130 GeV.

\section{Search of Higgs$\rightarrow\tau\tau\rightarrow {\rm lepton} + \tau_{\rm jet}$ with 1 fb$^{-1}$}

\begin{figure}[b]
\begin{center}
\includegraphics[width=.27\textwidth]{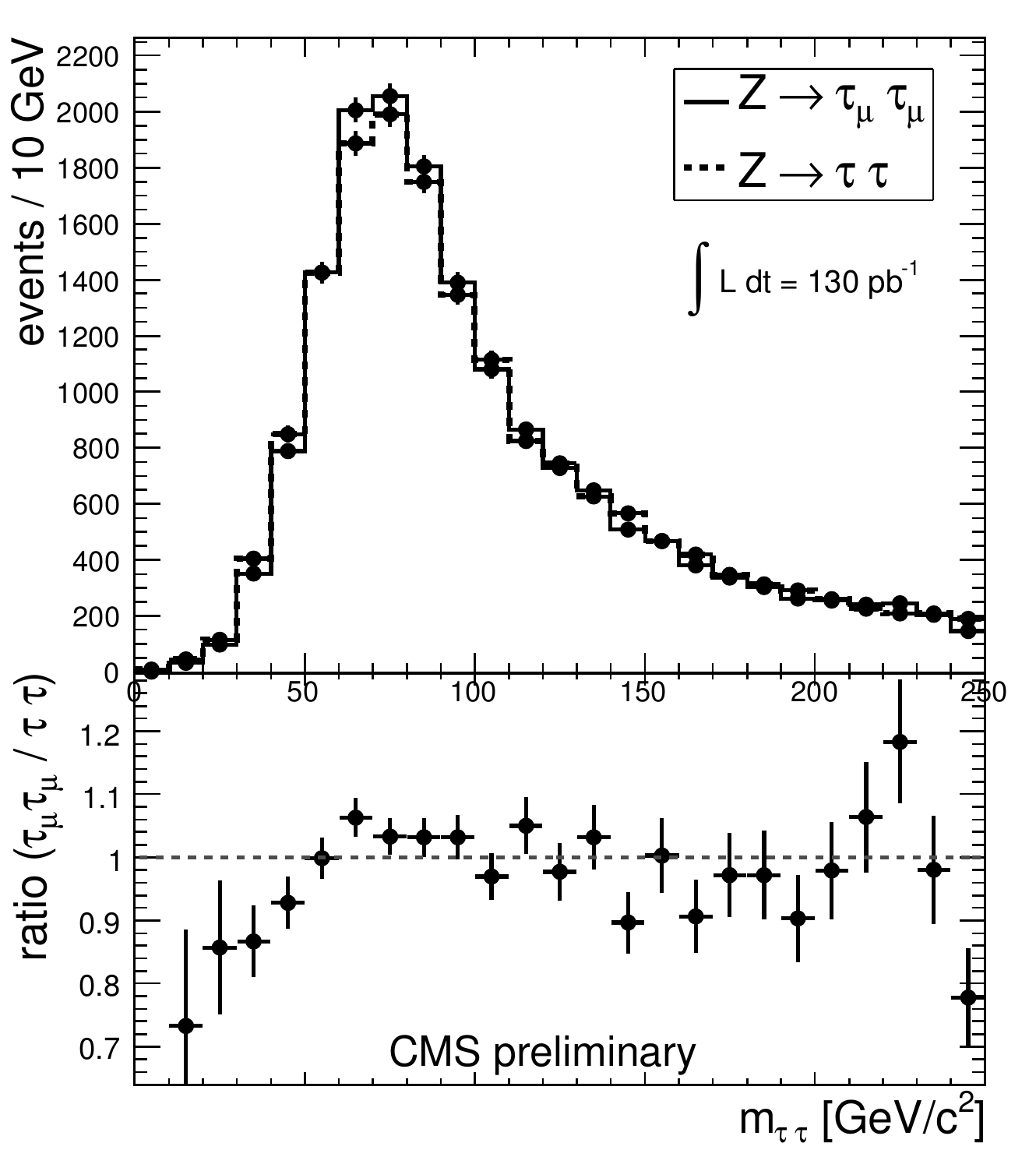}
\includegraphics[width=.30\textwidth]{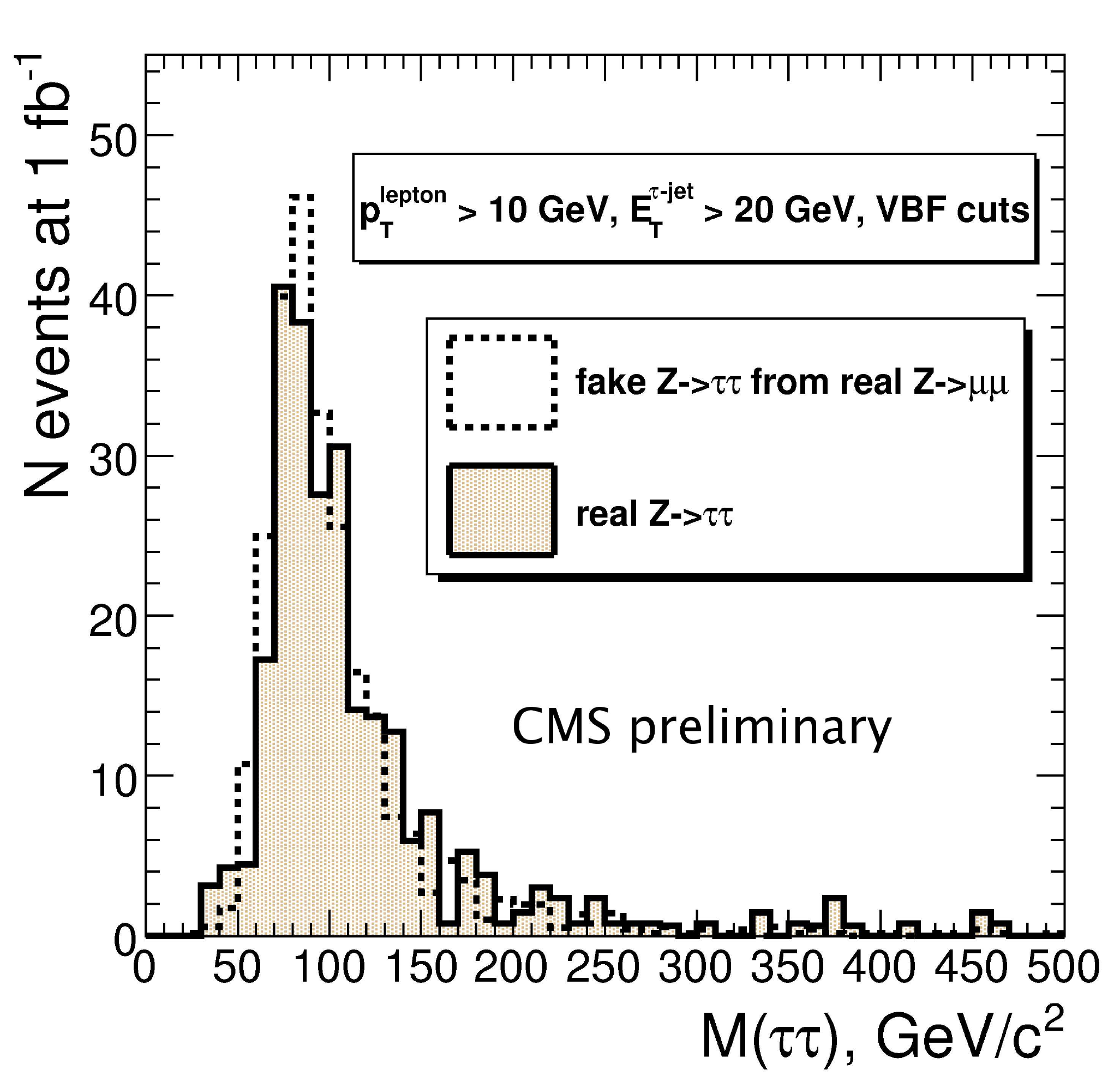}
\caption{Reconstructed di-$\tau$ mass for real and fake $Z\rightarrow\tau\tau$ events for the 
final states (left) $\tau\tau\rightarrow\mu\nu\nu+\mu\nu\nu$ from
inclusive Drell-Yan events and (right) $\tau\tau\rightarrow l\nu\nu+\tau_{jet}\nu$ from $Z$+jets events.}
\label{fig:zmass}
\end{center}
\end{figure}

\vspace*{-.2cm}
A selection strategy for the search of VBF Higgs $\rightarrow\tau\tau\rightarrow {\rm lepton} + \tau_{\rm jet}$ 
with 1 fb$^{-1}$ has been developed and is described in detail in~\cite{PAS_HIG_08_008}. The di-$\tau$
invariant mass will be analyzed to search for the presence of a Higgs boson in the region above the $Z\rightarrow
\tau\tau$ mass peak. It is important to know well the shape of the $Z\rightarrow\tau\tau$ background.
The dominant uncertainty comes from the modeling of the missing transverse momentum related
to the effects of pile-up, underlying event and the calorimeter noise and response. A method to
model  the di-$\tau$ mass has been developed~\cite{PAS_HIG_08_001}. $Z\rightarrow\mu\mu$ data events are 
selected and the muons are removed from the real event. Di-$\tau$ Monte Carlo events are generated with the same
kinematics as the real muons and their detector response is fully simulated. Finally the real $Z\rightarrow\mu\mu$
events with the muons removed and the simulated di-$\tau$ events are super-imposed to form one event, $Z\rightarrow\tau_{\mu}\tau_{\mu}$,
and the di-$\tau$ mass is calculated. The reconstructed di-$\tau$ mass for real and fake $Z\rightarrow\tau\tau$ events  for inclusive Drell-Yan and  $Z$+jets events are shown in Fig.~\ref{fig:zmass}. A good agreement between the
di-$\tau$ mass shapes is obtained.

\begin{figure}[t]
\begin{center}
\includegraphics[width=.29\textwidth]{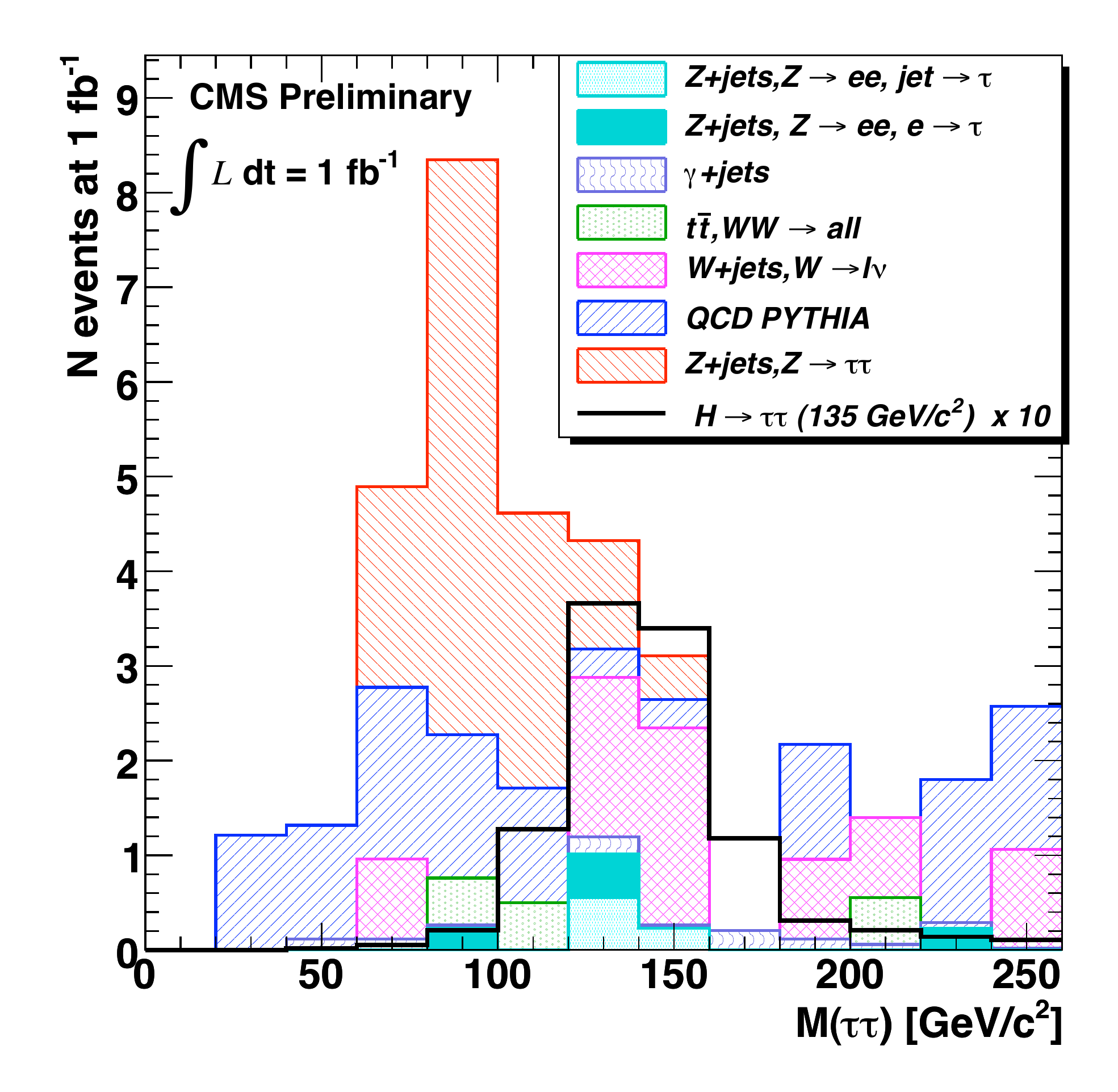}
\includegraphics[width=.4\textwidth]{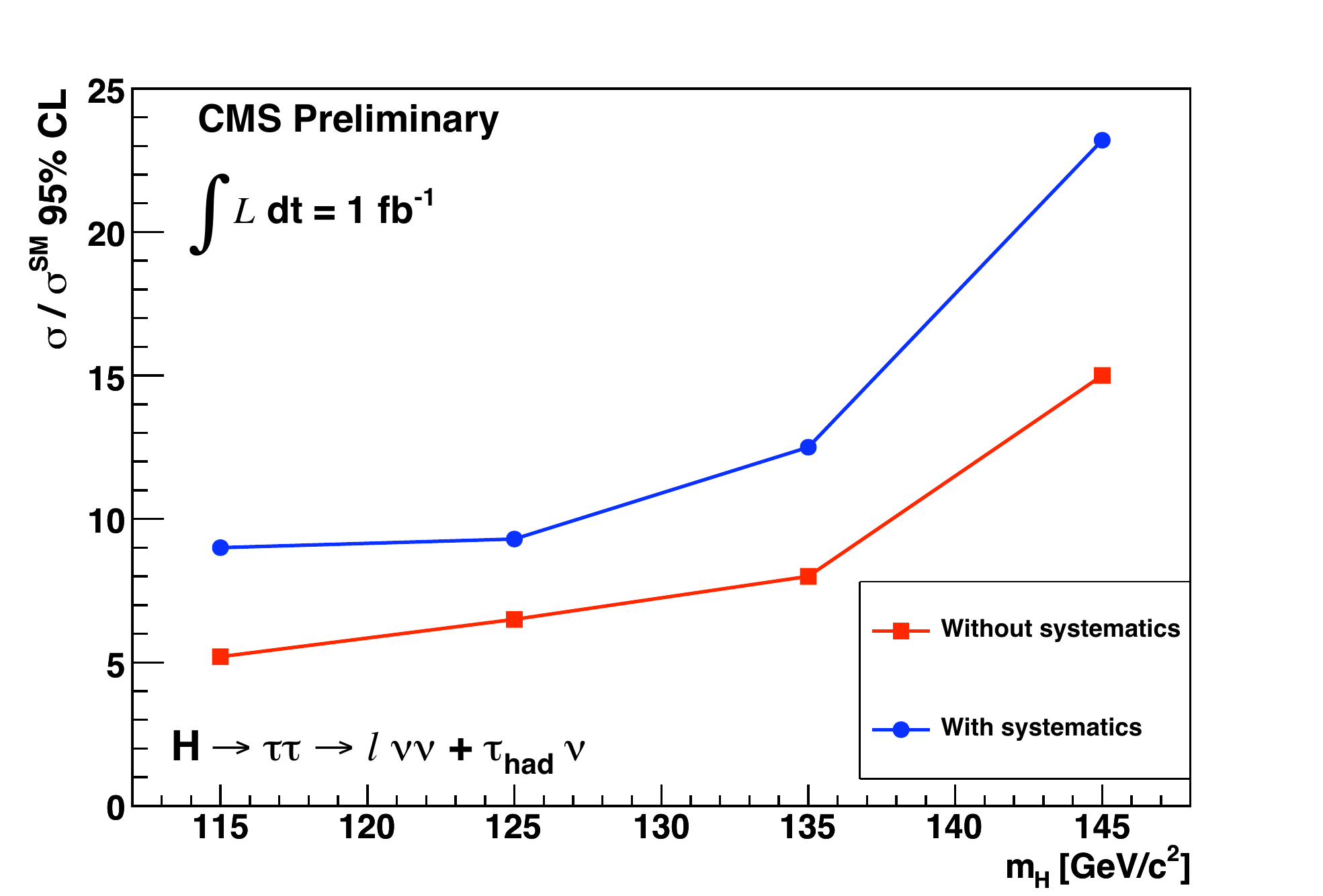}
\caption{Di-$\tau$ mass distribution of expected backgrounds with 1 fb$^{-1}$ after all selection. 
Backgrounds are shown cumulative. The signal mass distribution scaled by a factor 10 is also shown 
for $M_{H}$= 135 GeV.}
\label{fig:htautau}
\end{center}
\end{figure}

The expected di-$\tau$ mass distribution for the background and the Higgs signal for 1 fb$^{-1}$ is shown in Fig.~\ref{fig:htautau} (left).
A profile likelihood method is used to evaluate the upper limit on the number of signal events. Figure~\ref{fig:htautau} (right) 
shows the expected 95\% $CL$ limit on the cross section times branching ratio as a function
of the Higgs boson mass.   

\section{Conclusion}
A selection strategy for the Standard Model Higgs boson produced in vector boson fusion decaying to a pair of $\tau$ leptons with 1 fb$^{-1}$ of early CMS data at the LHC has been presented. No signal evidence is expected and upper limit on the cross section times branching ratio is evaluated. Prospective analyses for the $H \rightarrow \gamma\gamma$, $WW$ and $\tau\tau$ decay channels for a luminosity of 30\ fb$^{-1}$ have also been discussed. For a Higgs boson mass in the range 115 to 140 GeV, an observation with a significance above 2 standard deviations is expected in the H to $\gamma\gamma$ channel, and above 3 standard deviations in the H to $\tau\tau$ channel. The H to WW channel offers a discovery reach above 5 sigma in the mass range of 140 to 200 GeV.

\end{document}